\documentclass[runningheads]{llncs}

\usepackage[usenames,dvipsnames]{xcolor}
\usepackage{tikz}
\usepackage{wrapfig}
\usepackage{subfigure}
\usepackage{fixltx2e}
\usepackage{amsmath}
\usepackage{amssymb}
\usepackage{amsfonts}
\usepackage{latexsym}
\usepackage{xspace}
\usepackage{rotating}
\usepackage{enumerate}
\usepackage{paralist}
\usepackage{pgfplots}
\usepackage{booktabs}
\usepackage{enumitem}
\usepackage{etoolbox}

\usepackage{float}
\usepackage{pifont}

\usetikzlibrary{patterns,arrows,calc,shapes,shadows,decorations.pathmorphing,decorations.pathreplacing,automata,shapes.multipart,positioning,shapes.geometric,fit,circuits,trees,shapes.gates.logic.US,fit}

\usetikzlibrary{automata,fit}
\usetikzlibrary{decorations.pathmorphing}
\tikzstyle{every picture}=[thick]
\tikzstyle{every loop}=[->]
\tikzstyle{every scope}=[>=latex]
\tikzstyle{dot}=[circle,thick,minimum size=0.5mm,fill=black, inner sep=1pt]
\tikzstyle{every state}=[draw=black,line width=.5pt,fill=white,minimum size=10pt,initial text=]

\pgfplotscreateplotcyclelist{mycolor}{%
blue,every mark/.append style={fill=blue!80!black},mark=*, densely dashed\\%
violet,every mark/.append style={fill=violet!80!black},mark=square*, densely dashed\\%
BurntOrange!60!black,every mark/.append style={fill=BurntOrange!80!black},mark=otimes*\\%
red,mark=diamond*,every mark/.append style={fill=red!80!black}\\%
}

\input{dft-tikz}

\newcommand{\AND}{\textsf{AND}\xspace}
\newcommand{\OR}{\textsf{OR}\xspace}
\newcommand{\PAND}{\textsf{PAND}\xspace}
\newcommand{\POR}{\textsf{POR}\xspace}
\newcommand{\VOT}[1]{\textsf{VOT}\textsubscript{#1}\xspace}
\newcommand{\SEQ}{\textsf{SEQ}\xspace}
\newcommand{\SPARE}{\textsf{SPARE}\xspace}

\newcommand{\PDEP}{\textsf{PDEP}\xspace}
\newcommand{\BE}{\textsf{BE}\xspace}

\newcommand{\MTTF}{\ensuremath{\textsf{MTTF}}}
\newcommand{\VTTF}{\ensuremath{\textsf{VTTF}}}

\newcommand{\cmark}{\ding{51}}%
\newcommand{\xmark}{\ding{55}}%

\definecolor{lightblue}{RGB}{224,224,255}
\definecolor{lightred}{RGB}{255,224,224}
\definecolor{lightgreen}{RGB}{224,255,224}
\definecolor{lightyellow}{RGB}{255,255,224}
\definecolor{lightpurple}{RGB}{255,224,255}
\definecolor{darkerred}{RGB}{64,0,0}
\definecolor{darkred}{RGB}{128,0,0}
\definecolor{darkblue}{RGB}{0,0,128}
\definecolor{darkgreen}{RGB}{0,128,0}
\definecolor{darkpurple}{RGB}{128,0,128}

\makeatletter
\renewcommand{\paragraph}{%
  \@startsection{paragraph}{4}%
  {\z@ }{-5\p@ \@plus -4\p@ \@minus -4\p@ }{-0.5em \@plus -0.22em \@minus -0.1em}%
  {\normalfont\normalsize\itshape}%
}
\makeatother

\title{Advancing Dynamic Fault Tree Analysis}
\subtitle{Get succinct state spaces fast and synthesise failure rates}

\author{Matthias Volk, Sebastian Junges, Joost-Pieter Katoen}
\institute{Software Modeling and Verification, RWTH Aachen University, Germany}

\titlerunning{Advancing Dynamic Fault Tree Analysis}
\authorrunning{Matthias Volk, Sebastian Junges, Joost-Pieter Katoen}

\renewcommand{\subsubsection}[1]{\smallskip\noindent\textbf{#1.}}
\newcommand{\dftscale}{0.6}

\begin{document}
\maketitle

\begin{abstract}
This paper presents a new state space generation approach for dynamic fault trees (DFTs) together with a technique to synthesise failures rates in DFTs. 
Our state space generation technique aggressively exploits the DFT structure --- detecting symmetries, spurious non-determinism, and don't cares. 
Benchmarks show a gain of more than two orders of magnitude in terms of state space generation and analysis time.
Our approach supports DFTs with symbolic failure rates and is complemented by parameter synthesis.
This enables determining the maximal tolerable failure rate of a system component while ensuring that the mean time of failure stays below a threshold.
\end{abstract}

\section{Introduction}
\label{sec:intro}

Fault tree analysis is a prominent technique in reliability engineering.
Dynamic fault trees (DFTs)~\cite{DBB90,handbook2002} are an expressive model catering for common dependability patterns, such as spare management, functional dependencies, and sequencing. 
The \emph{state space generation} process is one of the main bottlenecks in DFT analysis.
DFT analysis mainly focuses on the mean time to failure --- what is the expected time of the failure? --- and reliability --- how likely is the system operational up to time $t$? 
These analyses require DFTs where all component failure rates are known.
In practice, this rarely holds.
A practically relevant question thus is to \emph{synthesise} the component failure rates ensuring a given mean time.

This paper presents three main advances to state-of-the-art DFT analysis: (1) fast generation of succinct state spaces, (2) the analysis of several measures-of-interest that go beyond mean time and reliability, and (3) the synthesis of (possibly partially) unknown failure rates in DFTs for mean time and more.

\paragraph{Fast generation of succinct state spaces.}
Our approach is a modern version of one of the first DFT semantics~\cite{CSD00} as used in the \texttt{Galileo} tool~\cite{SDC99} that caters for possible \emph{non-determinism}, as in~\cite{Boudali2010}.
To obtain succinct state spaces, we tailor two successful techniques from the field of model checking --- symmetry reduction~\cite{CEJS98} and partial-order reduction~\cite[Ch. 8]{Baier2008} --- to DFTs, and combine this with don't care detection.
We \emph{aggressively exploit the DFT structure}: detect symmetries, i.e., isomorphic sub-DFTs and stochastic independencies while pruning sub-DFTs that become obsolete (don't care) after the occurrence of some faults.
This is combined with \emph{detecting superfluous non-determinism} such that certain failure orderings are irrelevant yielding a simpler and cheaper analysis.

\paragraph{Beyond reliability and availability.}
By exploiting powerful state-of-the-art quantitative model checking techniques~\cite[Ch.\ 10]{Baier2008} we support a broad range of measures-of-interest.
This includes reliability and mean time to failure (MTTF), the probability to reach a certain DFT configuration e.g., where certain subDFTs have failed and others have not, conditional MTTF --- what is the MTTF given that certain DFT elements failed? --- and the variance of the time to failure.

\paragraph{Failure rate synthesis.}
We support DFTs whose failure rates are (possibly partially) unknown.
These unknown (or: symbolic) rates are represented by parameters, or functions thereof; e.g., components may fail with rate $\lambda$, $2 \lambda$, etc., where $\lambda$ is unknown.
Our slim state space generation techniques support symbolic rates.
We complement this by a sound and complete technique to \emph{synthesise} all values of symbolic rates that ensure the MTTF (and various other measures) to be below a given threshold.
To the best of our knowledge, this is the first failure rate synthesis technique for DFTs.
In addition, the \emph{sensitivity} of the MTTF on the symbolic rates can be determined, as in alternative techniques \cite{OD00}.
 
\paragraph{Experimentation.}
We have realised a prototypical implementation of the aforementioned techniques.
In addition to the original DFT elements in \texttt{Galileo}, we support probabilistic dependencies~\cite{MPBC06}, nested spares~\cite{Boudali2010} and priority or-gates~\cite{Walker2009}.
Experiments have been conducted on all benchmark DFTs from~\cite{JGKRS15}; a rich collection of DFTs gathered from the literature and from industrial case studies.
The experiments reveal that our slim state space generation technique significantly outperforms the best competitor for DFTs, the tool \texttt{DFTCalc}~\cite{ABBGS13}.
For a majority of the benchmarks, our approach yields a speed-up of two to four orders of magnitude.
Failure rate synthesis works for the moderately-sized models in the literature (up to 20 basic events) with up to three unknown rates.

\section{Dynamic fault trees}
\label{sec:dfts}
\begin{figure}[tb]
\subfigure[\BE]{
 \centering
\makebox[0.065\linewidth]{
\scalebox{0.62}{
 \begin{tikzpicture}[  scale=.8,font=\LARGE,text=black, every node/.style={transform shape}, node distance=0.3cm]
	\node[be] (pand) {};
	\node[above=of pand] (output) {};
	\draw[-] (pand) -- (output);
\end{tikzpicture}}}
 \label{Fig:BE} 
}
\subfigure[\small$\VOT k$]{
 \centering 
\makebox[0.095\linewidth]{
\scalebox{0.62}{
 \begin{tikzpicture}   
    \node[and3] (and) {\rotatebox{270}{$k$}};
    \node[below=0.4 cm of and.input 1, xshift=-0.2cm]  (i1) {};
    \node[below=0.3 cm of and.input 2]  (dots) {$\hdots$};
    
    \node[below=0.4 cm of and.input 3, xshift=0.2cm] (i2) {};
    
    \draw[-] (and.input 1) -- (i1);
    \draw[-] (and.input 3) -- (i2);
  \end{tikzpicture}}}
 \label{Fig:VOT}
}
   \subfigure[$\OR$]{
  \centering
\scalebox{0.62}{
 \begin{tikzpicture}   
    \node[or3] (and) {};
    \node[below=0.4 cm of and.input 1, xshift=-0.2cm]  (i1) {};
    \node[below=0.3 cm of and.input 2]  (dots) {$\hdots$};
    
    \node[below=0.4 cm of and.input 3, xshift=0.2cm] (i2) {};
    
    \draw[-] (and.input 1) -- (i1);
    \draw[-] (and.input 3) -- (i2);
  \end{tikzpicture}
  }
 \label{Fig:OR} 
}
 \subfigure[$\AND$]{
  \centering
\makebox[0.085\linewidth]{
\scalebox{0.62}{
 \begin{tikzpicture}   
    \node[and3] (and) {};
    \node[below=0.4 cm of and.input 1, xshift=-0.2cm]  (i1) {};
    \node[below=0.3 cm of and.input 2]  (dots) {$\hdots$};
    
    \node[below=0.4 cm of and.input 3, xshift=0.2cm] (i2) {};
    
    \draw[-] (and.input 1) -- (i1);
    \draw[-] (and.input 3) -- (i2);
  \end{tikzpicture}
  }
  }
 \label{Fig:AND} 
}
 \subfigure[\PAND]{
 \centering
\makebox[0.1\linewidth]{
 \scalebox{0.62}{
   \begin{tikzpicture}   
    \node[and3] (and) {};
    \node[triangle,scale=1.62,yshift=-3.5,xscale=0.80] (triangle_a) at (and) {};
    \node[below=0.4 cm of and.input 1, xshift=-0.2cm]  (i1) {};
    \node[below=0.3 cm of and.input 2]  (dots) {$\hdots$};
    
    \node[below=0.4 cm of and.input 3, xshift=0.2cm] (i2) {};
    
    \draw[-] (and.input 1) -- (i1);
    \draw[-] (and.input 3) -- (i2);
  \end{tikzpicture}}}
  \label{Fig:DFTElements_PAND}
  
 }
\subfigure[\POR]{
  \centering
\makebox[0.083\linewidth]{
  \scalebox{0.62}{
    \begin{tikzpicture}   
    \node[or2] (and) {};
    \node[btriangle,scale=1.61,yscale=0.915, xshift=-0.113cm] (triangle_b) at (and) {};
    \node[below=0.4 cm of and.input 1]  (i1) {};PANDs are commonly included i
    
    \node[below=0.4 cm of and.input 2] (i2) {};
    
    \draw[-] (and.input 1) -- (i1);
    \draw[-] (and.input 2) -- (i2);
  \end{tikzpicture}}}
  \label{Fig:DFTElements_POR}
 }
\subfigure[\SPARE]{
\centering
\makebox[0.115\linewidth]{
\scalebox{0.62}{
  \begin{tikzpicture}   
    \node[spare] (and) {};
    \node[below=0.4 cm of and.P]  (i1) {};

    \node[below=0.4 cm of and.SA] (i2) {};
    \node[below=0.3 cm of and.SC] (i3) {$\hdots$};
    
    \node[below=0.4 cm of and.SE, xshift=0.3cm] (i4) {};
    
    \draw[-] (and.P) -- (i1);
    \draw[-] (and.SA) -- (i2);
    \draw[-] (and.SE) -- (i4);
  \end{tikzpicture} }}
  \label{Fig:DFTElements_SPARE}
 }
 \subfigure[\SEQ]{
  \centering
  \makebox[0.083\linewidth]{
  \scalebox{0.62}{
    \begin{tikzpicture}   
    \node[seq] (and) {$\rightarrow$};
    \node[below=0.4 cm of and.250, xshift=-0.2cm]  (i1) {};
    \node[below=0.3 cm of and.270]  (dots) {$\hdots$};
    
    \node[below=0.4 cm of and.290, xshift=0.2cm] (i2) {};
    
    \draw[-] (and.250) -- (i1);
    \draw[-] (and.290) -- (i2);
  \end{tikzpicture}}}
  \label{Fig:DFTElements_SEQ}
 }
\subfigure[\PDEP]{
   \centering
   \scalebox{0.62}{
      \begin{tikzpicture}   
    
    \node[fdep] (and) {};
    \node[above=0.07cm of and.center] (x) {$p$};
    \node[below=0.7 cm of and.T, xshift=-0.2cm]  (i1) {};
    
    \node[below=0.4 cm of and.EB] (i2) {};
    \draw[-] (and.T) -- (i1);
    \draw[-] (and.EB) -- (i2);
    
  \end{tikzpicture}
  }
    \label{Fig:DFTElements_PDEP}
 }
 \caption{Node types in ((a)-(d)) static and (all) dynamic fault trees.}
 \label{Fig:DFTElements}
\end{figure}
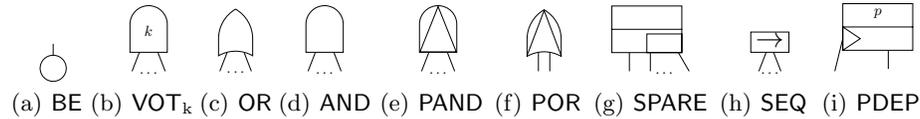

Fault trees (FTs) are directed acyclic graphs with typed nodes. 
The leaves, i.e., nodes without successors (or: \emph{children}), are \emph{basic events} (\BE{}s). 
All other nodes are \emph{gates}. 
The \emph{top event} (or: root) is a specifically identified node. 
An FT fails, if its top event fails. 
%
%
For the sake of simplicity, we assume that \BE{}s represent component failures. 
Initially, every \BE is \emph{operational}; it \emph{fails} if the event occurs. 
A gate fails if its \emph{failure condition} over its children is fulfilled. 
The key gate for static fault trees (SFTs) is the \emph{voting} gate (denoted \VOT{$k$}) with \emph{threshold} $k$. 
The failure condition for a node $x$ of type \VOT{$k$} is given by ''$x$ fails, if $k$ of its children have failed''. 
A \VOT{$1$} gate equals an OR-gate, while a \VOT{$k$} with $k$ children equals an AND-gate. 
These gates are shown in Fig.~\ref{Fig:DFTElements}(b)-(d). 
\subsection{Dynamic nodes}
To overcome the limitations \cite{JGKS16} of SFTs, several extensions commonly referred to as \emph{Dynamic Fault Trees} (DFTs) have been introduced. 
A main feature of these extensions is that they feature an internal state, e.g., the order in which events fail influences the internal state, and thus whether the top event has failed.
The extensions introduce several new node types; we categorise them as \emph{priority gates}, \emph{dependencies}, \emph{restrictions}, and \emph{spare gates}.

\subsubsection{Priority gates}
Priority gates extend static gates by imposing a condition on the ordering of failing children. 
A \emph{priority-and} (\PAND) node fails if all its children have failed in the order from left to right. 
Fig.~\ref{Fig:PandVsSeq_A} depicts a \PAND with children $A$ and $B$. 
It fails if $A$ fails first and then (or simultaneously) $B$ fails. 	
If $B$ fails first, the \PAND becomes \emph{fail-safe}.
The \emph{priority-or} (\POR) node~\cite{Walker2009} only fails if the left-most child fails before any of the other children does. 
Priority-gates allow for order dependent failure propagation.

\subsubsection{Dependencies}
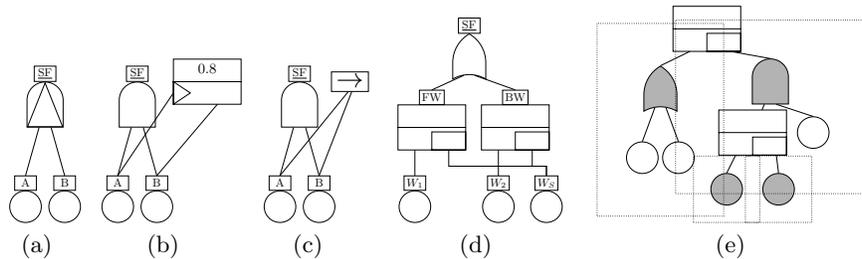
\begin{figure}[b]
\centering
\subfigure[]{ \centering
  \makebox[.1\linewidth]{ \scalebox{\dftscale}{\begin{tikzpicture}[scale=.6,text=black]
\centering
	\node[and2] (pand_a)  {};
	\node[triangle,scale=1.62,yshift=-3.5,xscale=0.80] (triangle_a) at (pand_a) {};
	\node[labelbox] (panda_label) at (pand_a.east){\underline{SF}};
	
	\node[be, below=1.4cm of pand_a.input 1, xshift=-0.3cm] (c1) {};
	\node[labelbox] (c1_label) at (c1.north){A};
	
	\node[be, below=1.4cm of pand_a.input 2, xshift=0.3cm] (cn) {};
	\node[labelbox] (cn_label) at (cn.north){B};
	
	\draw[-] (pand_a.input 1)   -- (c1_label.north);
	\draw[-] (pand_a.input 2)   -- (cn_label.north);
   
  \end{tikzpicture}}}
   \label{Fig:PandVsSeq_A}
 }
 \subfigure[]{ \centering
  \makebox[.14\linewidth]{ \scalebox{\dftscale}{\begin{tikzpicture}[scale=.6,text=black]
\centering
	\node[and2] (pand_a) {};
	\node[labelbox] (panda_label) at (pand_a.east){\underline{SF}};
	
	 \node[fdep, right=0.8cm of pand_a] (fdep) {};
    \node[above=0.07cm of fdep.center] (x) {$0.8$};
	\node[be, below=1.4cm of pand_a.input 1, xshift=-0.3cm] (c1) {};
	\node[labelbox] (c1_label) at (c1.north){A};
	
	\node[be, below=1.4cm of pand_a.input 2, xshift=0.3cm] (cn) {};
	\node[labelbox] (cn_label) at (cn.north){B};
	
	\draw[-] (pand_a.input 1)   -- (c1_label.north);
	\draw[-] (pand_a.input 2)   -- (cn_label.north);
	\draw[-] (fdep.T) -- (c1_label.north);
	\draw[-] (fdep.EB) -- (cn_label.north);
  \end{tikzpicture}}}
 \label{Fig:PDEPexample}
}
\subfigure[]{ \centering
  \makebox[.14\linewidth]{ \scalebox{\dftscale}{\begin{tikzpicture}[scale=.6,text=black]
\centering
	\node[and2] (pand_a) {};
	\node[labelbox] (panda_label) at (pand_a.east){\underline{SF}};
	
	\node[seq,right=0.7cm of pand_a] (seq) {$\rightarrow$}; 
	
	\node[be, below=1.4cm of pand_a.input 1, xshift=-0.3cm] (c1) {};
	\node[labelbox] (c1_label) at (c1.north){A};
	
	\node[be, below=1.4cm of pand_a.input 2, xshift=0.3cm] (cn) {};
	\node[labelbox] (cn_label) at (cn.north){B};
	
	\draw[-] (pand_a.input 1)   -- (c1_label.north);
	\draw[-] (pand_a.input 2)   -- (cn_label.north);
	\draw[-] (seq.260) -- (c1_label.north);
	\draw[-] (seq.280) -- (cn_label.north);
  \end{tikzpicture}}}
 \label{Fig:PandVsSeq_B}
}
\subfigure[]{
\scalebox{\dftscale}{
 \begin{tikzpicture}[scale=.6,text=black, node distance=1cm]
\centering
	\node[or2] (and) {};
	\node[labelbox] (andl) at (and.east) {\underline{SF}};
	\node[spare,below=of and,fill=white!100!black, xshift=-1.2cm] (spare1) {};
	
	\node[spare,right=0.35cm of spare1,fill=white!100!black] (spare2) {};
	\node[labelbox] (rl1) at (spare1.north) {FW};
	\node[labelbox] (rl2) at (spare2.north) {BW};
	
	\node[be,node distance=0.9cm,  below =of spare1.P] (p1) {};
	\node[be,node distance=0.9cm,  below =of spare2.P] (p2) {};
	\node[be,right =0.35cm of p2] (S) {};
	\node[labelbox] (pl1) at (p1.north) {$W_1$};
	\node[labelbox] (pl2) at (p2.north) {$W_2$};
	\node[labelbox] (sl) at (S.north) {$W_S$};

	\draw[-] (rl1.north) -- (and.input 1);
	\draw[-] 	(rl2.north) -- (and.input 2);
	\draw[-] 	(spare1.P) -- (pl1.north);
        \draw[-] 	(spare2.P) -- (pl2.north);
        \draw[-] (spare1.SC) -| +(0,-0.6)  -| (sl.north);
        \draw[-] (spare2.SC) -| +(0,-0.6)  -| (sl.north);
\end{tikzpicture}
}
\label{Fig:SpareExample}
}
\subfigure[]{
	\scalebox{\dftscale}{
\begin{tikzpicture}[scale=.6,text=black, node distance=1.1cm, box/.style={draw, densely dotted, inner sep=4pt}]]
   \centering
 \node[spare,fill=white!100!black, xshift=-3.8cm] (spare1) {};
 \node[or2,fill=black!30!white, below=1.4cm of spare1.center, yshift=1.4cm] (a1) {};
 
 \node[be, below=of a1.center, xshift=-0.4cm] (a2) {};
 \node[be, below=of a1.center, xshift=0.4cm] (a3) {};
 
 \node[and2,fill=black!30!white,below=1.2cm of spare1.center, yshift=-1cm] (b1) {};
 
 \node[spare,below=0.6cm of b1, xshift=-0.8cm] (spare2) {};
 \node[be, below=of b1, anchor=east, xshift=0.9cm] (b2) {};
 
  \node[be,fill=black!30!white, below=0.4cm of spare2.P, xshift=-0.2cm] (c1) {};
  \node[be,fill=black!30!white, below=0.4cm of spare2.SC, xshift=0.2cm] (d1) {}; 
  
  \node[box, fit=(a1) (a2) (a3)] {};
  \node[box, fit=(b1) (spare2) (b2)] {};
  \node[box, fit=(c1)] {};
  \node[box, fit=(d1)] {};
  
  \draw[-] (spare1.P) -- (a1.output);
  \draw[-] (spare1.SD) -- (b1.output);
  \draw[-] (a1.input 1) -- (a2.north);
  \draw[-] (a1.input 2) -- (a3.north);
  \draw[-] (b1.input 1) -- (spare2);
  \draw[-] (b1.input 2) -- (b2.north);
  \draw[-] (spare2.P) -- (c1.north);
  \draw[-] (spare2.SD) -- (d1.north);

 \end{tikzpicture}
 }
  \label{fig:sparemodules}	
}
\caption{Simple examples of dynamic nodes.}	
\label{fig:examples}
\end{figure}

Dependencies do not propagate a fault to their parents but are triggered by their first child. 
Upon triggering, they affect some BEs, the dependent events. 
We consider \emph{probabilistic dependencies} (PDEPs) \cite{MPBC06}. 
Once the trigger of a PDEP fails, its dependent events fail with probability $p$.
Fig.~\ref{Fig:PDEPexample} shows a PDEP where the failure of trigger $A$ causes a failure of BE $B$ with probability $0.8$ (provided it has not failed before).
\emph{Functional dependencies} (FDEPs) are PDEPs with probability one. 

\subsubsection{Restrictions}
Restrictions do not propagate failures but rather limit possible failure propagations. 
\emph{Sequence enforcers} (\SEQ{}s) assure that their children only fail from left to right. 
(This differs from priority-gates that do not prevent certain orderings, but propagate if an ordering is met.)
The DFT in Fig.~\ref{Fig:PandVsSeq_B} fails if $A$ and $B$ have failed (in any order) but the SEQ enforces that $A$ fails prior to $B$. 
This DFT is never fail-safe.

\subsubsection{Spare gates}
Spare-gates (\SPARE{}s) are the most complex gates in DFTs. 
Consider the DFT in Fig.\ref{Fig:SpareExample} modelling (part of) a motor bike with a spare wheel. 
If either wheel fails, the motor bike fails. 
Both wheels can be replaced by the spare wheel but not both. 
The spare wheel is less likely to fail as long as it isn't used (warm). 
Assume the front wheel fails. 
The spare wheel is available and used, and its failure rate is increased (hot). 
If any other wheel fails, then no spare wheels are available anymore, and the \SPARE and the DFT fails.

\SPARE{}s have a child they use. 
If this child fails, the \SPARE tries to use a spare child (left to right) --- a process we call \emph{claiming}. 
Only operational children that are not used by another \SPARE can be claimed. 
If claiming fails, the \SPARE fails. 
This behaviour is extended by an \emph{activation mechanism}.
As in~\cite{Boudali2010}, \SPARE{}s may have (independent) subDFTs as children. 
This includes nested \SPARE{}s. 
A \emph{spare module} is a set of nodes linked to a child of a \SPARE via a path without an intermediate \SPARE. 
Every leaf of a spare module is either a \BE or a \SPARE. 
Each child of a \SPARE thus represents a spare module, cf.\ Fig.~\ref{fig:sparemodules} where boxes are spare modules and shaded nodes are the representatives.
\SPARE{}s which are not nested are \emph{active}. 
For each active \SPARE, all nodes in the spare module of the used child are also active. 
\BE{}s which are active fail with their active failure rate, \BE{}s which are passive fail with their passive failure rate (warm events) or cannot fail (cold events).
More details can be found in~\cite{JGKS16}.
 
\subsection{Syntactic restrictions}
We are rather liberal w.r.t. dynamic gates, but have to impose syntactic restrictions as in \cite{ABBGS13} to exclude DFTs with undefined behaviour.
These restrictions are:
\begin{inparaenum}[(a)]
	\item \VOT{$k$} have at least $k$ children;
	\item the top level event is a gate or a BE;
	\item\label{FDEP-no-parent} \PDEP{}s and restrictions have no parents;
	\item all dependent events are \BE{}s;
	\item\label{Spare-independent} spare modules, i.e., subDFTs under a \SPARE, do not overlap;
	\item\label{PrimSpace-no-sharing} primary spare modules are not shared between \SPARE{}s.
\end{inparaenum}

\section{State space generation}
\label{sec:statespace}
The goal for our state space generation is to produce a Markov model which is subject to further analysis. As operational model, we use \emph{Markov Automata}.

\subsection{Markov Automata}

Markov Automata (MA)  \cite{EHZ10b} extend continuous-time Markov chains (CTMCs) with non-determinism. MA are state transition systems whose transitions between states are either labeled with rates (i.e., non-negative real numbers), or with actions.  
The former transitions specify a random delay and correspond to the failures in DFTs; the latter are used to select the handling of a triggered PDEP.
Delay transitions relate a source state with a target state; action transitions relate a state to a probability distribution over states.
An action transition thus yields a new state with a given likelihood.
MA are a slight variant of the operational model for DFTs used in~\cite{Boudali2010}; they differ in allowing discrete probabilistic branching which are used to model \PDEP{}s.
We introduce MAs by example.

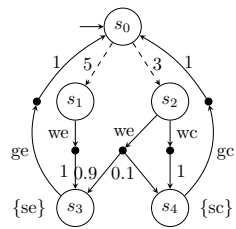
\begin{wrapfigure}{r}{0.25\textwidth}
\scalebox{0.7}{
\centering
 \begin{tikzpicture}
  \node[state,initial, initial text=] (s0) {$s_0$};
  \node[state,below=0.7cm of s0, xshift=-0.9cm] (s1) {$s_1$};
  \node[state,below=0.7cm of s0, xshift=0.9cm] (s2) {$s_2$};
  \node[state, below=1.3cm of s1] (s3) {$s_3$};
  \node[state, below=1.3cm of s2] (s4) {$s_4$};
  \node[circle, draw, fill=black, scale=0.4,below=0.5cm of s1] (i1) {};
  \node[circle, draw, fill=black, scale=0.4,] (i2) at (i1 -| s0) {};
  \node[circle, draw, fill=black, scale=0.4, below=0.5cm of s2] (i3) {};
  \node[circle, draw, fill=black, scale=0.4, left=0.3cm of s1] (i4) {};
  \node[circle, draw, fill=black, scale=0.4, right=0.3cm of s2] (i5) {};
  \node[left=0.07cm of s3] (l3) {$\{ \text{se} \}$};
  \node[right=0.07cm of s4] (l4) {$\{ \text{sc} \}$};

  \draw[->, dashed] (s0) -- node[left]{$5$} (s1);
  \draw[->, dashed] (s0) -- node[right]{$3$}(s2);
  \draw[->] (s1) -- node[left] {we} (i1);
  \draw[->] (s2) -- node[left] {we} (i2);
  \draw[->] (s2) -- node[right] {wc} (i3);
  \draw[->] (i1) -- node[left] {$1$}(s3);
  \draw[->] (i2) -- node[left] {$0.9$} (s3);
  \draw[->] (i2) -- node[left] {$0.1$} (s4);
  \draw[->] (i3) -- node[right] {$1$}(s4);
  \draw[->] (s3) edge[bend left=30] node[left] {ge} (i4);
  \draw[->] (s4) edge[bend right=30] node[right] {gc} (i5);
  \draw[->] (i4) edge[bend left=14] node[left] {$1$}(s0);
  \draw[->] (i5) edge[bend right=14] node[right] {$1$}(s0);
  
 \end{tikzpicture}
 }
 \vspace{-4mm}
 \caption{Example MA.} 
 \vspace{-3mm}
 \label{Fig:ExMA}
\end{wrapfigure}

Fig.~\ref{Fig:ExMA} shows an MA for a coffee machine, used by inhabitants of room A (IoA) and B (IoB). 
IoA (IoB) arrive at the machine at a rate of $5$ IoA/hour ($3$ IoB/hour). 
They can either have coffee or espresso. 
All IoA want espresso (action \textsf{we}), while IoB non-deterministically want coffee (action \textsf{wc}) or espresso. 
IoB wanting espresso are with probability $0.1$ too sleepy and select coffee. 
Users always get their selected product (\textsf{ge}, \textsf{gc}).
In state $s_0$, either an IoA or an IoB arrives at the machine (evolving into $s_1$, $s_2$). 
In state $s_1$ espresso is selected, whereas in $s_2$ a choice between actions \textsf{we} and \textsf{wc} is made.
Selecting \textsf{we} in $s_2$ results in $s_3$ with probability $0.1$ and in $s_4$ with probability $0.9$.
The user then gets the product and the automaton returns to initial $s_0$. 
For simplicity, the products' preparation time is not modelled.
%

\subsection{State space generation}
As in \texttt{Galileo}, we construct a \emph{fault tree automaton} (FTAut) from a DFT. 
We then translate the FTAut to an MA, which we further simplify and analyse.
The FTAut consists of states and labelled transitions. 

\subsubsection{States}
We give each node in the DFT a unique id. 
A state in the FTAut is a mapping from ids to its status: \emph{operational} (OP), \emph{failed} (F), \emph{fail-safe} (FS), or \emph{don't care} (X). Additionally,  we store the \emph{currently used child} (CUC) of operational \SPARE{}s and for spare module representatives their \emph{activity}, i.e.\ whether the module is active (A) or passive (P).
We initialise all nodes as operational, the CUCs and activate modules as described in Section \ref{sec:dfts}.

\subsubsection{Transitions}
State changes originate from the failure of \BE{}s. 
As the probability of two rate-governed \BE{}s to fail simultaneously is zero, \BE{}s never fail simultaneously. 
When considering dependencies, this assumption no longer has to hold. 
To avoid problems with causalities as described in \cite{JGKS16}, and to directly resolve spare races \cite{JGKS16}, we assume that dependent events fail immediately after the triggering \BE. 
W.l.o.g.\ we assume that \PDEP{}s have a single dependent event.

Given a source state and an operational \BE $x$ that fails, we copy the source state and additionally mark $x$ with F, and compute the target state.
In a bottom-up fashion, we iterate over the gates. For each gate, we check the failure condition. If the failure condition holds, we mark the gate as failed. If a CUC of a \SPARE fails, we iterate over its remaining children and check whether they are not listed as the CUC of any of their parents and whether they are still operational. If so, we update the CUC, otherwise, we mark the \SPARE as failed. 
We iterate over all restrictions, and check whether any of their failure conditions hold; if so, we skip the transition at hand.
We then reiterate over all gates, and check if the fail-safe condition holds (i.e.\ if it cannot fail in the future), we mark the gate FS.
We then iterate top-down over all nodes. If all parents of a node are either failed or fail-safe, we mark the node as don't care (DC-propagation).  

\begin{figure}[t] 
\subfigure[FTAut]{
\scalebox{\dftscale}{
\begin{tikzpicture}[baseline=(and.north), scale=.6,text=black]
\tikzstyle{state} = [rectangle,draw=black,line width=.5pt,fill=white,minimum size=10pt,initial text=]
\centering
    \node[state,initial, initial text=] (s0) {(OP, OP, OP)};
    \node[state,below=of s0, xshift=-1.1cm] (s1) {(F, OP, OP)};
    \node[state,below=of s0, xshift=1.1cm] (s2) {(X, X, FS)}; 
    \node[state, below=of s1] (s3) {(X, X, F)};
  
    \draw[->] (s0) -- node[above left] {A} (s1);
    \draw[->] (s0) -- node[above right] {B} (s2);
    \draw[->] (s1) -- node[left] {B} (s3);
\end{tikzpicture}
\label{Fig:ExploreExample}
}}
\subfigure[MA]{
\scalebox{\dftscale}{
\begin{tikzpicture}[baseline=(and.north), scale=.6,text=black]
    \node[state,initial, initial text=] (s0) {$s_0$};
    \node[state,below=0.85cm of s0, xshift=-0.5cm] (s1) {$s_1$};
    \node[state,below=0.85cm of s0, xshift=0.5cm] (s2) {$s_2$}; 
    \node[state, below=0.85cm of s1] (s3) {$s_F$};
  
    \draw[->, dashed] (s0) -- node[above left] {1} (s1);
    \draw[->, dashed] (s0) -- node[above right] {2} (s2);
    \draw[->, dashed] (s1) -- node[left] {2} (s3);
\end{tikzpicture}
\label{Fig:MAExample}
}}
\subfigure[State construction example]{
\centering
\scalebox{\dftscale}{
\begin{tikzpicture}[baseline=(and.north), scale=.6,text=black]
\centering
    \node[and4] (and) {};
	\node[labelbox] (and_label) at (and.east) {\underline{$J$}};
	\node[and2,below=1.6cm of and, yshift=0.8cm] (or2) {};
	\node[triangle,scale=1.62,yshift=-3.5,xscale=0.80] (triangle_a) at (or2) {};
	\node[labelbox] (g1_label) at (or2.output) {$I$};
	
	\node[or2,below=1.6cm of and, yshift=1.9cm] (or1) {};
	\node[labelbox] (g2_label) at (or1.output) {$H$};
	 \node[fdep, left=0.8cm of or1, yshift=0.4cm] (fdep) {};
    \node[above=0.07cm of fdep.center] (x) {$0.8$};
    \node[labelbox] (fdep_label) at (fdep.north) {$G$};
	\node[seq,right=2.5cm of and, yshift=-2cm] (seq) {$\rightarrow$};

	\node[be,below=2.7cm of and.center, xshift=-1.1cm] (A) {};
	\node[labelbox] (a_label) at (A.north) {$B$};
	\node[be,left=0.3cm of A] (B) {};
	\node[labelbox] (b_label) at (B.north) {$A$};
	\node[be,right=0.3cm of A] (C) {};
	\node[labelbox] (c_label) at (C.north) {$C$};
	\node[be,right=0.3cm of C] (D) {};
	\node[labelbox] (d_label) at (D.north) {$D$};
	\node[be,right=0.3cm of D] (E) {};
	\node[labelbox] (e_label) at (E.north) {$E$};
	\node[be,right=0.3cm of E] (F) {};
	\node[labelbox] (f_label) at (F.north) {$F$};
	
	\draw[-] (and.input 1) -- (g2_label.north);
	\draw[-] (or1.input 2) -- (a_label.north);
	\draw[-] (or1.input 1) -- (b_label.north);
	\draw[-] (and.input 2) -- (g1_label.north);
	\draw[-] (or2.input 1) -- (c_label.north);
	\draw[-] (or2.input 2) -- (d_label.north);
	\draw[-] (and.input 3) -- (e_label.north);
	\draw[-] (and.input 4) -- (f_label.north);
	\draw[-] (seq.270) -- (e_label.north);
	\draw[-] (seq.290) -- (f_label.north);
	
	\draw[-] (fdep.T) -| +(-0.4, -0.8) -- (b_label.north);
	\draw[-] (fdep.EB) -| +(0, -0.2) -- (c_label.north);
	
\end{tikzpicture}}
\label{Fig:StateConstructionExample}
}
\subfigure[Symmetry]{
\scalebox{\dftscale}{
\begin{tikzpicture}[baseline=(and.north), scale=.6,text=black]
\centering
    \node[and2] (and) {};
	\node[labelbox] (and_label) at (and.east) {\underline{$\text{PC}$}};
	\node[and2,below=1.6cm of and] (or2) {};
	\node[triangle,scale=1.62,yshift=-3.5,xscale=0.80] (triangle_a) at (or2) {};
	\node[labelbox] (g1_label) at (or2.output) {$A'$};
	
	\node[and2,below=1.6cm of and, yshift=1.5cm] (or1) {};
	\node[triangle,scale=1.62,yshift=-3.5,xscale=0.80] (triangle_b) at (or1) {};
	\node[labelbox] (g2_label) at (or1.output) {$A$};

	\node[be,below=2.7cm of and.center, xshift=-0.5cm] (A) {};
	\node[labelbox] (a_label) at (A.north) {$C$};
	\node[be,left=0.3cm of A] (B) {};
	\node[labelbox] (b_label) at (B.north) {$B$};
	\node[be,right=0.3cm of A] (C) {};
	\node[labelbox] (c_label) at (C.north) {$B'$};
	\node[be,right=0.3cm of C] (D) {};
	\node[labelbox] (d_label) at (D.north) {$C'$};
	
	\draw[-] (and.input 1) -- (g2_label.north);
	\draw[-] (or1.input 2) -- (a_label.north);
	\draw[-] (or1.input 1) -- (b_label.north);
	\draw[-] (and.input 2) -- (g1_label.north);
	\draw[-] (or2.input 1) -- (c_label.north);
	\draw[-] (or2.input 2) -- (d_label.north);
\end{tikzpicture}
\label{Fig:SymRedExample}
}}
\vspace{-1mm}
\caption{Dedicated examples.}
\vspace{-2mm}
\end{figure}
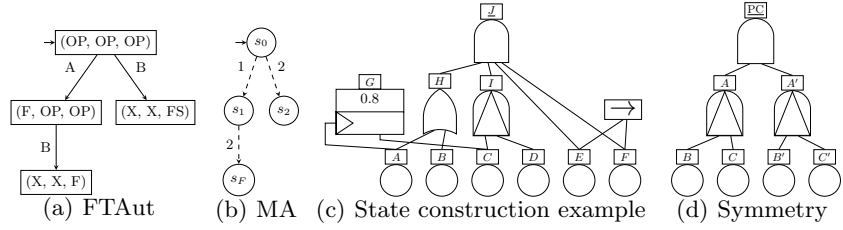

\begin{example}
The FTAut of the DFT in Fig.~\ref{Fig:PandVsSeq_A} is given in Fig.~\ref{Fig:ExploreExample}.
Initially, all nodes are operational.
If $B$ initially fails, the \PAND becomes fail-safe, and thus $A$ and $B$ both become don't care.
The resulting state is (X, X, FS).
If $A$ however initially fails, $B$ and the \PAND remain operational.
An additional failure of $B$ then causes the top event to fail.
DC-propagation yields the state (X, X, F).

Now consider Fig.~\ref{Fig:StateConstructionExample}. 
Initially, every node is operational. 
$A$'s failure causes $H$ to fail and makes $B$ don't care. 
This yields a transition from the initial state to state (F, X, OP, OP, OP, OP, OP, F, OP, OP).
In this state, the PDEP is triggered, yielding a state (with probability $0.8$) in which $C$ failed, and the same state (with probability $0.2$) as $C$ does not fail.
A failure of $D$ in the initial state does not trigger a failure of the \PAND $I$; in fact $I$ becomes fail-safe, and this is propagated to $J$, i.e., DC-propagation marks all children (and their children) X.
This together yields a transition from the initial state to a state in which all nodes are marked X, except for $G$.
Finally, from the initial state, propagating a failure of node $F$ is discarded as the restriction fails (by $F$ failing before $E$.) 
 
The initial state for nodes $(W_1, W_2, W_S, \text{FW}, \text{BW}, \text{SF})$ in Fig.~\ref{Fig:SpareExample} is (OP, OP, OP, $W_1$, $W_2$, OP) where $W_1$, $W_2$ are the CUCs and as initially the CUCs are active, the activity for $W_1, W_2, W_S$ is given as (A, A, P). 
A failure of $W_1$ is propagated to FW. 
As its CUC fails, it checks further children. 
$W_S$ is operational and not a CUC, therefore, the resulting state is (F, OP, OP, $W_S$, $W_2$, OP) and (A, A, A). 
From that state, $W_2$'s failure yields (F, F, OP, $W_S$, F, F) after failure propagation, as the only remained child of BW is already claimed. 
DC-propagation yields the state (F, X, OP, $W_S$, X, F) and (A, A, A).
\end{example}

As rate-governed transitions have probability $0$ to fire at time $0$, we either have immediate transitions or rate transitions. Thus, for each state, we check if any \PDEP{}s are triggered. If so, we mark the state as \emph{immediate} and add two outgoing transitions for each triggered \PDEP: One where the \PDEP transmits the failure and one where it doesn't.   
Otherwise, we mark the state as \emph{Markovian}, and add transitions for each \BE which has (in the given state) a failure rate $\neq 0$. 

\subsubsection{Translation}
The translation from the FTAut to the MA is now straightforward.
The state spaces of the FTAut and the MA are equal. 
Each MA state is labeled with the status of the DFT nodes.
For Markovian states, each transition labelled with a \BE $x$ is translated into a delay transition with as rate the failure rate of $x$. 
For \BE{}s in passive spare modules, we take their passive failure rate.
Each immediate state has a non-deterministic choice over all triggered \PDEP{}s in the DFT. 
For each \PDEP, we get a probabilistic branching, where with probability $p$ the \PDEP propagates the failure, whereas with $1{-}p$ it does not.

\subsection{Optimisations}

\subsubsection{Technical aspects}
We use a selection of well-known techniques to reduce the overhead of propagation: The states are encoded as bit-vectors, and during exploration, we use an expanded state representation. 
By exploiting depth-first search, we keep the set of states that we explore later on small. 
Work lists keep only track of the nodes we need to consider. 
Overriding failed and fail-safe nodes by don't care, we can merge states which differ only in their past behaviour, but not in their future behaviour.
State spaces are reduced by bisimulation.

\subsubsection{Partial-order reduction}
In many DFTs, the actual order in which subsets of BEs fail is not crucial. We exploit this for dependencies, where --- instead of exploring all interleavings over the triggered events --- we aim to only explore a single order. We adapt a technique called (static) \emph{partial order reduction} \cite{Baier2008} to DFTs. 
Based on a static analysis, we identify which dependencies can be executed in arbitrary order, and expand only a canonical order.

\subsubsection{State elimination}
In MA, we can eliminate probabilistic branching by adopting a \emph{state elimination} technique as used in \cite{Daw04}. 
In particular, this allows us to reduce MA without non-deterministic branching to CTMCs, which can be analysed much faster as non-determinism is absent.

\subsubsection{Modularisation}
Modularisation has been proposed in \cite{GD97}. It identifies independent subtrees in the DFT, analyses them separately, and combines the obtained results to the final result. If applicable, it is extremely powerful.

\subsubsection{Symmetry reduction}
Many parts in DFTs are symmetric.  This can be exploited (cf.\ \cite{CEJS98}) as follows. Given that we successfully detect the symmetry, we can use the fact that a fault has an analogous effect in symmetric parts. Moreover for isolated symmetric parts, if the node identities are not used in the analysis and the parts are only connected to the remaining DFT via the same node, we can \emph{exchange} the states of the parts, and thus assume that a fault in a symmetric part happened in an equivalent DFT. In the DFT in Fig.~\ref{Fig:SymRedExample}, we find two symmetric parts (the subtrees of $A$ and $A'$), which are independent. If we are only interested in the top level, we can use the exchange technique. That is, if both symmetric parts are in equivalent states (e.g., the initial state) and $A'$ fails, we can assume that $A$ failed instead. Now, the two parts are not in an equivalent state. However, after the additional failure of $A'$, the two parts are in an equivalent state again.

\section{Measures of interest}
\label{sec:measures}
Once the state space of a DFT is generated (in the form of an MA), several quantitative measures-of-interest can be determined.

\subsubsection{Measures and importance factors}
Various measures are based on the \emph{reliability function}, the cdf for the probability of a failure after a given time $t$. 
Another prominent measure is the \emph{mean time to failure} (MTTF), the expected time until a system failure. 
The \emph{variance} of the time to failure (VTTF) is obtained by $\text{Var}(X) = E[X^2] - E[X]^2$ for random variable $X$, the time to failure. 
The \emph{probability of failure} considers the limit probability of the reliability function for $t$ to $\infty$. 
This is of interest as in DFTs not all events fail eventually, cf.\ the DFT in Fig.~\ref{Fig:PandVsSeq_A}.
These measures can be used for single events in the DFT.
They can also be used for Boolean combinations of failed and operational gates, such as e.g., the expected time to a DFT state where (only) events $A$ and $C$ have failed.  
Another measure-of-interest is the \emph{expected number of faults before the DFT fails}; if this is high, it indicates that are various possibilities to take countermeasures.
The \emph{Fussell-Vesely importance factor} is the probability that a \BE has failed when the DFT fails~\cite{RS15}. 
An exemplary \emph{criticality importance factor} is the probability that a BE causes the DFT to fail. 
To evaluate the measures above we use efficient algorithms to verify CTMCs~\cite{BHHK03} or --- if non-determinism remains --- MA~\cite{GHHKT14}.
\begin{table}[t]
\centering
\begin{tabular}{@{}lll|cc|ccc@{}}
symbol 		& name  							&  	& cond. 		& par.syn.  & mod. 		& dc.    & sym.red.\\ \midrule 
$R_F(t)$ 	& Reliability at $t$ 				&  	&  \xmark 	& \xmark 	& \cmark 	& \cmark & \cmark \\ 
$\Pr_F$ 	& Probability of failure   			&  	& \cmark  	& \cmark 	& \cmark 	& \cmark & \cmark \\
$\MTTF_F$ 	& Mean time to failure  				&  	& \cmark 	& \cmark	& \xmark	& \cmark & \cmark\\
$\VTTF_F$ 	& Variance of time to failure   		&  	& \cmark 	& \cmark 	& \xmark	& \cmark & \cmark\\\midrule
			& Expected faults before failure   	&  	& \cmark 	& \cmark	& \xmark	& \xmark & \cmark\\
			& FV importance factor 	& 	& \cmark	& \cmark	& \xmark    & \xmark & $*$ \\ 
			& Criticality importance factor 		& 	& \cmark	& \cmark	& \xmark  	& \cmark & $*$ \\\bottomrule
\end{tabular}
\caption{Supported measures and importance factors}
\vspace{-4mm}
\label{tab:measures}
\end{table}

\subsubsection{Conditional measures}
All measures (except for reliability) can be conditioned on the occurrence of events, cf.\ the first column of Table~\ref{tab:measures}.
For example, as the probability of failure does not always equal one, the MTTF is not always defined. 
In this case a more reasonable measure is the MTTF conditioned on the fact that the DFT indeed eventually fails.

\subsubsection{Measure preservation under optimisations}
Techniques such as modularisation, DC-propagation and symmetry reduction are not applicable to all measures.
Their robustness wrt.\ the measures is indicated in the last columns of Table~\ref{tab:measures}, where $*$ means support of a light version.
Modularisation is powerful if a partial state space suffices. 
This is e.g., the case if the measure is compositional, i.e, the measure can be obtained from its subDFTs' measures.
This holds e.g., for reliability but not for MTTF.
Symmetry reduction requires a lack of identity (of DFT nodes).
Thus only a light variant of symmetry reduction can be applied to some measures. 
In the conditional variant, the lack of identity is also not always given.

\section{Parameter Synthesis}
\label{sec:synthesis}
\subsubsection{Problem}
The analysis discussed so far has two drawbacks: It requires all failure rates in the DFT to be given and does not guarantee any robustness w.r.t.\ perturbations.
The latter has been addressed by \emph{sensitivity analysis} \cite{OD00}.
These deficiencies inspired us to treat \emph{symbolic} failure rates, i.e. DFTs where failure rates and propagation probabilities in \PDEP{}s are given as polynomials over a set of parameters (pDFTs). 
Our state space construction technique is largely unaffected by this.
Our focus is on the failure rate synthesis in DFTs for any measure in Table~\ref{tab:measures} except $R_F(t)$, i.e., \emph{determine all values (of the symbolic rates) such that the DFT satisfies a given desired threshold on a measure}.
For simplicity, we focus on DFTs that (after our reductions) obey no non-determinism, which applies to the vast majority of the DFTs in the literature.
Thus, the underlying state space of pDFTs can be reduced to a \emph{parametric} CTMC, i.e.\ a CTMC whose rates are polynomials over the DFT parameters.

\subsubsection{Approach}
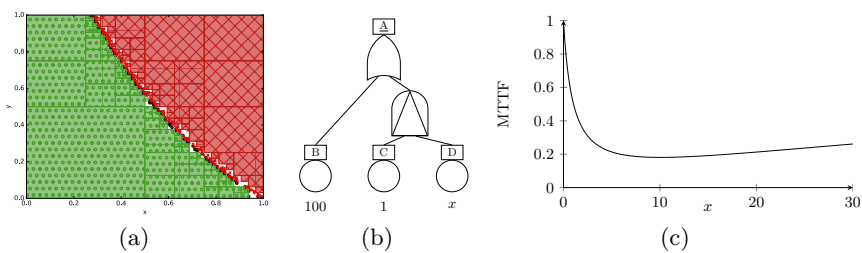
\begin{figure}[t]
\centering
\subfigure[]{
    \includegraphics[scale=0.20,trim=35 10 45 35,clip]{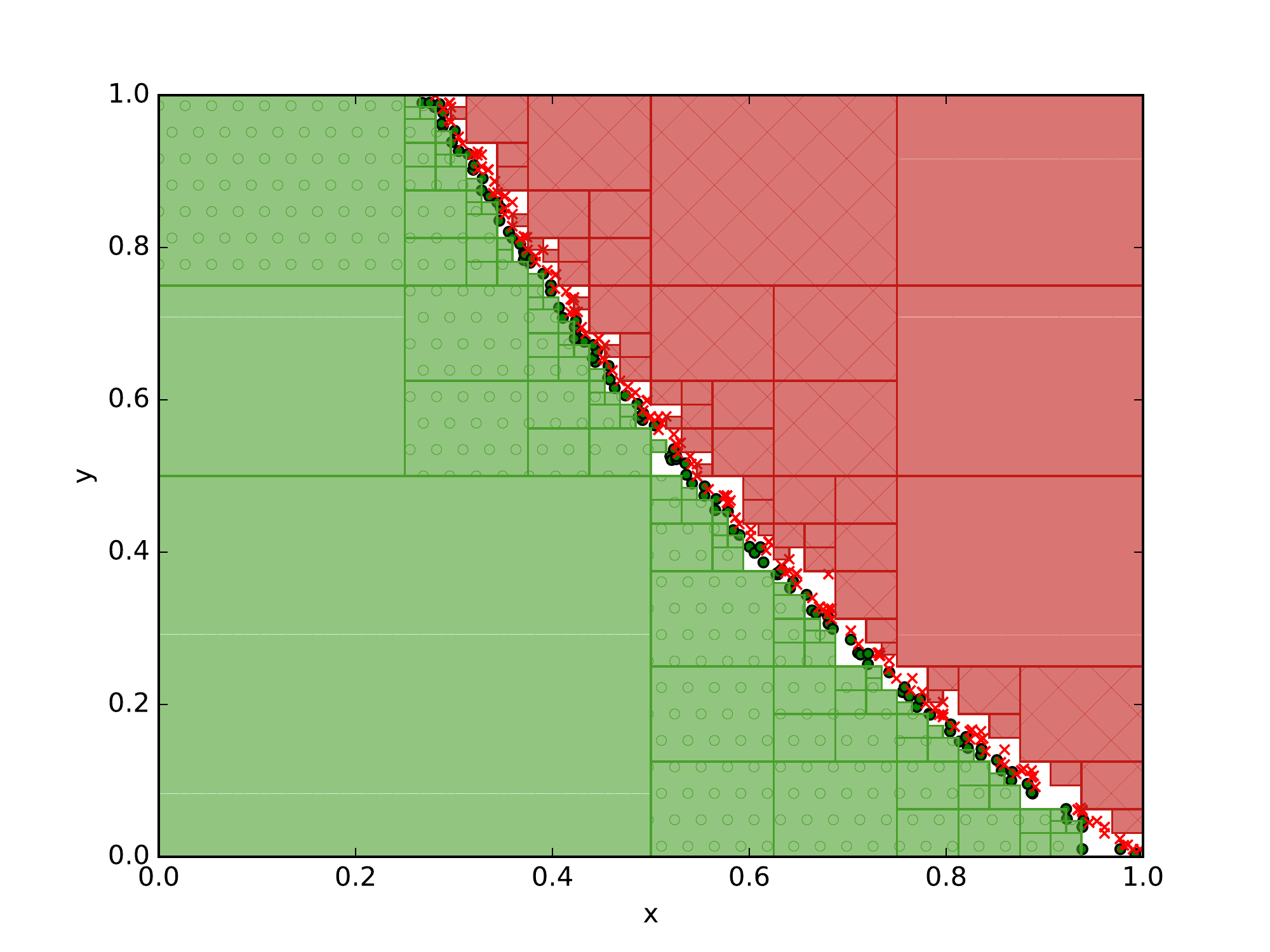}
\label{Fig:ProphesyOutput}
}
\subfigure[]{
	\scalebox{\dftscale}{
\begin{tikzpicture}[scale=.6,text=black]
\centering
    \node[or2] (and) {};
	\node[labelbox] (and_label) at (and.east) {\underline{A}};
	\node[and2,below=1.2cm of and, yshift=0.2cm] (or2) {};
	\node[triangle,scale=1.62,yshift=-3.5,xscale=0.80] (triangle_a) at (or2) {};
	\node[be,below=2.2cm of and.center] (C) {};
	\node[labelbox] (c_label) at (C.north) {C};
	\node[below=0.1cm of C] (cr_label) {$1$};
	\node[be,left=0.8cm of C] (B) {};
	\node[labelbox] (b_label) at (B.north) {B};
	\node[below=0.1cm of B] (br_label) {$100$};
	\node[be,right=0.8cm of C] (D) {};
	\node[labelbox] (d_label) at (D.north) {D};
	\node[below=0.1cm of D] (dr_label) {$x$};
	
	\draw[-] (and.input 1) -- (b_label.north);
	\draw[-] (and.input 2) -- (or2.output);
	\draw[-] (or2.input 1) -- (c_label.north);
	\draw[-] (or2.input 2) -- (d_label.north);
\end{tikzpicture}}
\label{Fig:NonMonoDFT}
}
\subfigure[]{
\begin{tikzpicture}[scale=0.65]
\begin{axis}[axis lines=left,x label style={at={(axis description cs:0.5,0.09)},anchor=north}, xlabel={$x$}, ylabel={MTTF}, width=7.5cm, height=5cm, ymin=0, ymax=1
]
\addplot[
black,
domain=0:30,
samples=201
] {(x^2+2*x+101)/((x+1) * (x+101))};
\end{axis}
\end{tikzpicture}

\label{Fig:NonMonoPlot}
}
\caption{(a) Sample output, (b) a sample parametric DFT, and (c) its MTTF.}	
\end{figure}

To enable the synthesis in pDFTs we exploit the parameter synthesis tool \texttt{PROPhESY}~\cite{dehnert-et-al-cav-2015}. Based on ideas in~\cite{Daw04}, it computes a closed form (precisely: a rational function) for a parametric CTMC and the measure of interest. 
To enable sensitivity analysis, it provides the derivative w.r.t.\ the parameters. 
On top of obtaining these functions, \texttt{PROPhESY} allows for parameter space partitioning --- using satisfiability-modulo-theory (SMT) techniques for non-linear arithmetic. 
That is, given a pDFT, we can synthesise for which parameter values the measure (e.g., MTTF) is above a threshold. 
An example output is depicted in Fig.~\ref{Fig:ProphesyOutput}. 
This plot was obtained for the DFT of Fig.~\ref{Fig:SpareExample} where $W_1$, $W_2$ and $W_S$ have failure rates $x$, $1$ and $y$ respectively for unknown $x, y$.
The green boxes represent areas in which \emph{all} failure rates of $W_1$ and $W_S$ give rise to an MTTF that exceeds $1.5$, while the red boxes guarantee \emph{all} rates yield an MTTF below $1.5$.
For the white areas, none of the above statements can be made.
Note that this output is extremely valuable as it provides information about many (in fact uncountably many) failure rate combinations for which the MTTF is below or above the threshold.
We like to point out that obtaining this information is far from trivial, and intrinsically more involved than analysing a DFT where all failure rates are given.
Consider the small example DFT from Fig.~\ref{Fig:NonMonoDFT}, where D has a symbolic failure rate. 
The MTTF of the DFT is given by the plot in Fig.~\ref{Fig:NonMonoPlot}. 
As the MTTF is not monotonic, the parameter synthesis is not straightforward.

\section{Experiments}
\label{sec:experiments}
\subsubsection{Set-up}
To evaluate the performance of our approach, we tested the performance of our tool on reliability and the MTTF assessment.  We compare with the state-of-the-art tool \texttt{DFTCalc}~\cite{ABBGS13} and assess the effect of our abstraction techniques. The experiments were conducted on an HP BL685C G7, 48 cores, 2.0GHz each, and 192GB of RAM. We restricted the RAM to 8GB and set a time-out of one hour for all experiments. 
We use the benchmark suite from  \cite{JGKRS15}. Besides the smaller HCAS and SAP sets, it contains the following benchmarks:\\
{\bf HECS.}
The \emph{Hypothetical Example Computer System (HECS)} stems from the NASA handbook on FTs~\cite{handbook2002}. It features a computer system consisting of a processor, a memory unit (MU) and an interface consisting of hard- and software. \\
{\bf MCS.}
The \emph{Multiprocessor Computing System (MCS)} contains computing modules consisting of a processor, a MU and two disks, the DFT was given in~\cite{MPBC06}.

{\bf RC.}
The \emph{Railway Crossing (RC)} is an industrial case modelling failures at level crossing~\cite{GKSLR14}. 
It fails whenever any of the sensor-sets, barriers or controller fails. \\
{\bf SF.} 
The \emph{Sensor Filter (SF)} benchmark is a DFT that is automatically generated from an AADL (Architecture Analysis \& Design Language) system model~\cite{Bozzano2011}.

We used the simplified DFTs as produced in \cite{JGKRS15}, as this is shown to be beneficial for \texttt{DFTCalc}. 
For each instance, we tested reliability for $t=100$ and the MTTF. 
Further features were tested on a range of $>100$ crafted instances.

\begin{figure}[tb]
\centering
\subfigure[run time (seconds)]{
\begin{minipage}{0.47\textwidth}
\centering
\begin{tikzpicture}
    \begin{axis}[
        width=5.7cm,
        height=5.7cm,
        xmin=0.01,
        ymin=0.01,
        ymax=12600,
        xmax=12600,
        xmode=log,
        ymode=log,
        axis x line=bottom,
        axis y line=left,
        x label style={at={(axis description cs:0.5,0.01)},anchor=north},
        y label style={at={(axis description cs:0.16,0.5)},anchor=south},
        xtick={1, 60, 600, 3600},
        xticklabels={1, 60, 600, 3600},
        extra x ticks = {7200, 12600},
        extra x tick labels = {TO,MO},
        extra x tick style = { grid = major },
        ytick={1, 60, 600, 3600}, 
        yticklabels={1, 60, 600, 3600},
        extra y ticks = { 7200, 12600 },
        extra y tick labels = {TO,MO},
        extra y tick style = { grid = major },
        xlabel=\texttt{SToRMDFT},
        ylabel=\texttt{DFTCalc},
        yticklabel style={font=\tiny},
        xticklabel style={rotate=290, anchor=west, font=\tiny},
        legend pos=south east, 
        legend style={font=\tiny}]
    \addplot[ 
        scatter,only marks,
        scatter/classes={
        hecs={mark=square*,blue,mark size=1.30},
        mcs={mark=triangle*,red,mark size=1.30},
        rc={mark=*,brown, mark size=1.30},
        sf={mark=diamond*,purple, mark size=1.30}
        },  
        scatter src=explicit symbolic]
        table [col sep=comma, x=StormDFT, y=DFTCalc, meta=Name]
        {time.csv};
    \legend{HECS,MCS,RC,SF}
    \addplot[no marks] coordinates
        {(0.01,0.01) (3600,3600) };
    \addplot[no marks, dashed] coordinates
        {(0.01,0.1) (360,3600) };
    \addplot[no marks, dashed] coordinates
        {(0.01,1) (36,3600) };
    \end{axis}
 \end{tikzpicture}
 \end{minipage}
 \label{Fig:ExpScatterTime}
 }
 \hfill
 \subfigure[memory footprint (MB)]{
 \begin{minipage}{0.47\textwidth}
 \centering
\begin{tikzpicture}
    \begin{axis}[
        width=5.7cm,
        height=5.7cm,
        xmin=1000,
        ymin=1000,
        ymax=16000000,
        xmax=16000000,
        xmode=log,
        ymode=log,
        axis x line=bottom,
        axis y line=left,
        x label style={at={(axis description cs:0.5,0.01)},anchor=north},
        y label style={at={(axis description cs:0.14,0.5)},anchor=south},
        xtick={1000, 10000, 100000, 1000000},
        xticklabels={1, 10, 100, 1000},
        extra x ticks = {10000000, 16000000},
        extra x tick labels = {TO,MO},
        extra x tick style = { grid = major },
        ytick={1000, 10000, 100000, 1000000},
        yticklabels={1, 10, 100, 1000},
        extra y ticks = {10000000, 16000000},
        extra y tick labels = {TO,MO},
        extra y tick style = { grid = major },
        xlabel=\texttt{SToRMDFT},
        ylabel=\texttt{DFTCalc},
        yticklabel style={font=\tiny},
        xticklabel style={rotate=290, anchor=west, font=\tiny},
        legend pos=south east,
        legend style={font=\tiny}]
    \addplot[
        scatter,only marks,
        scatter/classes={
        hecs={mark=square*,blue,mark size=1.30},
        mcs={mark=triangle*,red,mark size=1.30},
        rc={mark=*,brown, mark size=1.30},
        sf={mark=diamond*,purple, mark size=1.30}
        },
        scatter src=explicit symbolic]
        table [col sep=comma, x=StormDFT, y=DFTCalc, meta=Name]
        {memory.csv};
    \legend{HECS,MCS,RC,SF}
    \addplot[no marks] coordinates
        {(1,1) (6000000,6000000) };
    \addplot[no marks, dashed] coordinates
        {(1,10) (600000,6000000) };
    \end{axis}
 \end{tikzpicture}
 \end{minipage}
\label{Fig:ExpScatterMem}
 }
 \subfigure[max. \# states in MC]{
 \begin{minipage}{0.42\textwidth}
 \vspace*{-1mm}
  \centering
\begin{tikzpicture}
    \begin{axis}[
        width=5.5cm,
        height=5.5cm,
        xmin=1,
        ymin=1,
        ymax=120000000,
        xmax=120000000,
        xmode=log,
        ymode=log,
        axis x line=bottom,
        axis y line=left,
        x label style={at={(axis description cs:0.5,0.01)},anchor=north},
        y label style={at={(axis description cs:0.16,0.5)},anchor=south},
        xtick={1, 1000, 1000000},
        xticklabels={1, 10$^3$, 10$^6$},
        extra x ticks = {50000000, 120000000},
        extra x tick labels = {TO,MO},
        extra x tick style = { grid = major },
        ytick={1, 1000, 1000000},
        yticklabels={1, 10$^3$, 10$^6$},
        extra y ticks = {50000000, 120000000},
        extra y tick labels = {TO,MO},
        extra y tick style = { grid = major },
        xlabel=\texttt{SToRMDFT},
        ylabel=\texttt{DFTCalc},
        yticklabel style={font=\tiny},
        xticklabel style={rotate=290, anchor=west, font=\tiny},
        legend pos=south east,
        legend style={font=\tiny}]
    \addplot[
        scatter,only marks,
        scatter/classes={
        hecs={mark=square*,blue,mark size=1.30},
        mcs={mark=triangle*,red,mark size=1.30},
        rc={mark=*,brown, mark size=1.30},
        sf={mark=diamond*,purple, mark size=1.30}
        },
        scatter src=explicit symbolic]
        table [col sep=comma, x=StormDFT, y=DFTCalc, meta=Name]
        {max_states.csv};
    \legend{HECS,MCS,RC,SF}
    \addplot[no marks] coordinates
        {(1,1) (30000000,30000000) };
    \addplot[no marks, dashed] coordinates
        {(1,10) (3000000,30000000) };
    \end{axis}
 \end{tikzpicture} 
 \end{minipage}
 \label{Fig:ExpScatterFinal}
}
 \begin{minipage}{0.55\textwidth}
 \hfill
 \subtable[\#solved \& total run time (seconds)]{
\scriptsize{
\begin{tabular}{l|r|r|r|r|r|r|r|r}
          & \multicolumn{4}{c|}{reliability} & \multicolumn{4}{c}{MTTF} \\
          & \multicolumn{2}{c|}{\texttt{DFTCalc}} & \multicolumn{2}{c|}{\texttt{SToRMDFT}} & \multicolumn{2}{c|}{\texttt{DFTCalc}} & \multicolumn{2}{c}{\texttt{SToRMDFT}} \\
          & \# & Time & \# & Time & \# & Time & \# & Time \\
\hline 
HECS(42) & 38 & 2.8e4 & 42 & 3.1e0 & 36 & 2.6e4 & 40 & 7.0e3 \\
MCS(42) & 40 & 2.1e4 & 42 & 2.1e1 & 38 & 1.9e4 & 38 & 2.1e3 \\
RC(38) & 29 & 2.7e4 & 38 & 2.1e0 & 29 & 2.7e4 & 38 & 6.5e1 \\
SF(30) & 26 & 1.6e4 & 30 & 1.8e0 & 25 & 1.4e4 & 29 & 4.4e3 \\
CAS(8) & 8 & 1.3e3 & 8 & 3.6e-1 & 8 & 1.3e3 & 8 & 3.6e-1 \\
SAP(4)  & 4 & 3.6e2 & 4 & 3.0e-1 & 4 & 3.2e2 & 4 & 1.6e-1 \\

\end{tabular}
\label{Table:Results}
}
}

\subtable[optimisation run time (seconds)]{
\scriptsize{
\begin{tabular}{l|r|r|r|r|r|r|r|r|r}
\setlength{\tabcolsep}{1pt}
          & \multicolumn{5}{c|}{reliability} & \multicolumn{4}{c}{MTTF} \\
          & \multicolumn{1}{c|}{none} &  \multicolumn{1}{c|}{SR} & \multicolumn{1}{c|}{DC} & \multicolumn{1}{c|}{Mod} & \multicolumn{1}{c|}{all} & \multicolumn{1}{c|}{none} & \multicolumn{1}{c|}{SR} & \multicolumn{1}{c|}{DC} & \multicolumn{1}{c}{all}\\
\hline 
HC$_2$ & 30.3 & 15.6 & 1.1 & 0.05 & 0.04 & 29.9 & 15.5 & 1.2 & 0.61 \\ 
MC$_2$ & 337.8 & 46.0 & 1.1 & 0.05 & 0.05 & 334.0 & 45.6 & 1.1 & 0.21 \\
RC$_{10}$ & 53.6 & 0.1 & 53.5 & 0.20 & 0.05 & 53.9 & 0.1 & 53.6 & 0.07\\
SF$_2^6$ & 22.1 & 7.4 & 0.3 & 0.04 & 0.04 & 22.3 &7.4 & 0.2 & 0.08\\
\end{tabular}
\label{Tab:ResComp}
}
}
\end{minipage}
\vspace*{-1mm}
\caption{Overview of the experimental results on four different benchmark sets.}
\label{Fig:ExpScatter}
\end{figure}
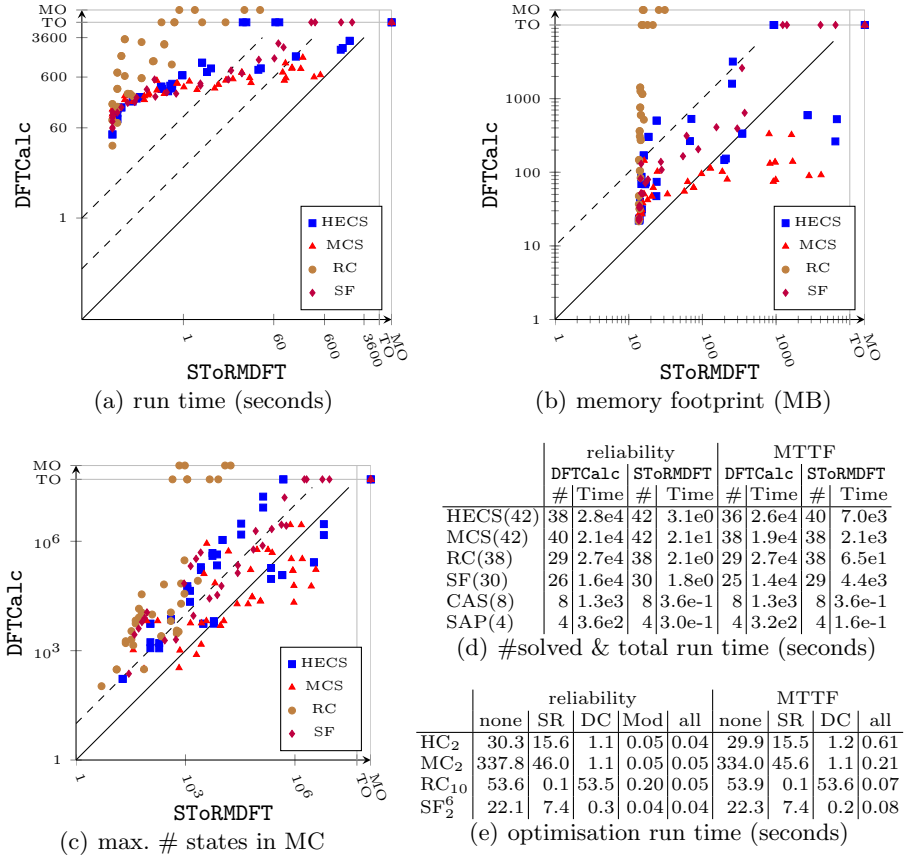

\subsubsection{Results}
Figures~\ref{Fig:ExpScatter}(a-c) compare the performance of our tool (referred to as \texttt{SToRMDFT}) with \texttt{DFTCalc} on MTTF (where modularisation is not applicable). 
All plots use a log-log-scale. 
Fig.~\ref{Fig:ExpScatterTime} presents the analysis time of a DFT.
This includes state space generation.
The lower dashed line indicates an advantage of our tool by a factor ten, the upper of a factor 100. 
The outer lines indicate TOs and MOs, respectively.
Fig.~\ref{Fig:ExpScatterMem} indicates the peak memory consumption as given by the operating system.  
Fig.~\ref{Fig:ExpScatterFinal} shows the peak intermediate state size. 
Table~\ref{Table:Results} summarises the performance on the benchmark sets --- it lists the number of benchmarks solved and the cumulative time needed for the solved benchmarks.
Table~\ref{Tab:ResComp} shows the effect of the individual optimisation  techniques (symmetry reduction, DC-propagation, modularisation) versus using all of them.

\subsubsection{Observations}
For non-parametric DFTs the performance is dominated by the state space construction. \texttt{SToRMDFT} creates intermediate state spaces that are often ten times smaller; especially for moderately-sized DFTs, this is done with a much lower overhead. 
This results in generating state spaces up to $5$ orders of magnitudes faster. 
The informed state space generation allows to stop exploring states where the measure of interest is settled. 
This advantage is best observed by comparing top events typed OR and AND.
The former requires significantly smaller state spaces, which is reflected by the smaller intermediate state spaces --- and leads to a significant advantage over \texttt{DFTCalc}. 
These effects are multiplied by aggressively applying symmetry reductions and DC-propagation. 
For many benchmarks, our abstractions directly yield the small bisimulation quotient. However, on some HECS and MCS instances, our symmetry reduction does not yet suffice and \texttt{DFTCalc} gains an advantage in terms of memory. 
Modularisation remains a powerful approach for assessing reliability. 
It profits additionally from the performance on small DFTs. 
Model-checking for reliability is for both \texttt{SToRMDFT} and \texttt{DFTCalc} so fast that our slightly better performance is hardly significant. For MTTF, \texttt{SToRMDFT} is significantly faster.

For parametric instances, the original DFTs from literature can be handled:
For, e.g., the standard HECS from literature it takes 5 seconds to compute the rational function with more than 400 terms in the numerator. Parameter synthesis for $90\%$ of the parameter space finishes within four minutes.  
However, scalability beyond these moderately-sized DFTs remains an open issue, as the parameters appear throughout the full state space. 

\section{Conclusions and future work}
\label{sec:conclusions}
We have presented a state space generation technique for DFTs that is more than two orders of magnitude faster than the state-of-the-art. 
  The technique is complemented with a new feature in DFT analysis --- the synthesis of failure rates for measures such as MTTF.
Future work includes the failure rate synthesis for reliability (e.g., using \cite{CDKP14}) and improve scalability for parameterised MTTF. 

\paragraph{Acknowledgement.}
We thank Christian Dehnert for fruitful discussions.
\bibliographystyle{splncs}
\bibliography{bibliography}

\begin{thebibliography}{10}

\bibitem{DBB90}
Dugan, J.B., Bavuso, S.J., Boyd, M.:
\newblock Fault trees and sequence dependencies.
\newblock In: Proc.\ of RAMS. (1990)  286--293

\bibitem{handbook2002}
Stamatelatos, M., Vesely, W., Dugan, J.B., Fragola, J., Minarick, J.,
  Railsback, J.:
\newblock Fault Tree Handbook with Aerospace Applications.
\newblock NASA Headquarters (2002)

\bibitem{CSD00}
Coppit, D., Sullivan, K.J., Dugan, J.B.:
\newblock Formal semantics of models for computational engineering: a case
  study on dynamic fault trees.
\newblock In: Proc. of ISSRE. (2000)  270--282

\bibitem{SDC99}
Sullivan, K., Dugan, J.B., Coppit, D.:
\newblock The {Galileo} fault tree analysis tool.
\newblock In: Proc.\ of FTCS. (1999)  232--235

\bibitem{Boudali2010}
Boudali, H., Crouzen, P., Stoelinga, M.I.A.:
\newblock A rigorous, compositional, and extensible framework for dynamic fault
  tree analysis.
\newblock IEEE Transactions on Dependable Secure Computing \textbf{7}(2) (2010)
   128--143

\bibitem{CEJS98}
Clarke, E.M., Emerson, E.A., Jha, S., Sistla, A.P.:
\newblock Symmetry reductions in model checking.
\newblock In: Proc.\ of CAV. Volume 6605 of LNCS, Springer (1998)  147--158

\bibitem{Baier2008}
Baier, C., Katoen, J.P.:
\newblock Principles of Model Checking.
\newblock MIT Press (2008)

\bibitem{OD00}
Ou, Y., Dugan, J.B.:
\newblock Sensitivity analysis of modular dynamic fault trees.
\newblock In: Proc.\ of IPDS. (2000)  35--43

\bibitem{MPBC06}
Montani, S., Portinale, L., Bobbio, A., Codetta-Raiteri, D.:
\newblock Automatically translating dynamic fault trees into dynamic {B}ayesian
  networks by means of a software tool.
\newblock In: Proc. of ARES. (2006)  6--

\bibitem{Walker2009}
Walker, M., Papadopoulos, Y.:
\newblock Qualitative temporal analysis: {T}owards a full implementation of the
  {F}ault {T}ree {H}andbook.
\newblock Control Engineering Practice \textbf{17}(10) (2009)  1115--1125

\bibitem{JGKRS15}
Junges, S., Guck, D., Katoen, J., Rensink, A., Stoelinga, M.:
\newblock Fault trees on a diet - automated reduction by graph rewriting.
\newblock In: Proc.\ of SETTA. Volume 9409 of LNCS, Springer (2015)  3--18

\bibitem{ABBGS13}
Arnold, F., Belinfante, A., van~der Berg, F., Guck, D., Stoelinga, M.:
\newblock Dftcalc: A tool for efficient fault tree analysis.
\newblock In: Proc.\ of SAFECOMP. Volume 8153 of LNCS.
\newblock Springer (2013)  293--301

\bibitem{JGKS16}
Junges, S., Guck, D., Katoen, J.P., Stoelinga, M.:
\newblock Uncovering dynamic fault trees.
\newblock In: Proc. of DSN. (2016) to appear.

\bibitem{EHZ10b}
Eisentraut, C., Hermanns, H., Zhang, L.:
\newblock On probabilistic automata in continuous time.
\newblock In: Proc.\ of LICS, IEEE Computer Society (2010)  342--351

\bibitem{Daw04}
Daws, C.:
\newblock Symbolic and parametric model checking of discrete-time {{M}arkov}
  chains.
\newblock In: Proc.\ of ICTAC. Volume 3407 of LNCS, Springer (2004)  280--294

\bibitem{GD97}
Gulati, R., Dugan, J.B.:
\newblock A modular approach for analyzing static and dynamic fault trees.
\newblock In: Proc.\ of RAMS. (1997)  57--63

\bibitem{RS15}
Ruijters, E., Stoelinga, M.I.A.:
\newblock Fault tree analysis: A survey of the state-of-the-art in modeling,
  analysis and tools.
\newblock Computer Science Review \textbf{15-16}(0) (2015)  29--62

\bibitem{BHHK03}
Baier, C., Haverkort, B.R., Hermanns, H., Katoen, J.:
\newblock Model-checking algorithms for continuous-time markov chains.
\newblock IEEE Trans.\ Softw.\ Eng. \textbf{29}(6) (2003)  524--541

\bibitem{GHHKT13}
{Guck}, D., {Hatefi}, H., {Hermanns}, H., {Katoen}, J.P., {Timmer}, M.:
\newblock Modelling, reduction and analysis of markov automata.
\newblock In: Proc.\ of QEST. Volume 8054 of LNCS, Berlin, Springer (2013)
  55--71

\bibitem{dehnert-et-al-cav-2015}
Dehnert, C., Junges, S., Jansen, N., Corzilius, F., Volk, M., Bruintjes, H.,
  Katoen, J.P., {\'{A}}brah{\'{a}}m, E.:
\newblock Prophesy: {A} probabilistic parameter synthesis tool.
\newblock In: Proc.\ of CAV. Volume 9206 of LNCS, Springer (2015)  214--231

\bibitem{GKSLR14}
{Guck}, D., {Katoen}, J.P., {Stoelinga}, M., {Luiten}, T., {Romijn}, J.:
\newblock Smart railroad maintenance engineering with stochastic model
  checking.
\newblock In: Proc.\ of RAILWAYS. Volume 104 of Civil-Comp Proceedings,
  Civil-Comp Press (2014)  299--314

\bibitem{Bozzano2011}
Bozzano, M., Cimatti, A., Katoen, J.P., Nguyen, V.Y., Noll, T., Roveri, M.:
\newblock Safety, dependability and performance analysis of extended {AADL}
  models.
\newblock The Computer Journal \textbf{54} (2011)  754--775

\bibitem{CDKP14}
Ceska, M., Dannenberg, F., Kwiatkowska, M.Z., Paoletti, N.:
\newblock Precise parameter synthesis for stochastic biochemical systems.
\newblock In: Proc.\ of CMSB. Volume 8859 of LNCS, Springer (2014)  86--98

\bibitem{DBLP:conf/tacas/BartocciGKRS11}
Bartocci, E., Grosu, R., Katsaros, P., Ramakrishnan, C.R., Smolka, S.A.:
\newblock Model repair for probabilistic systems.
\newblock In: Proc.\ of TACAS. Volume 6605 of LNCS, Springer (2011)  326--340

\end{thebibliography}

\end{document}